\documentstyle[epsf]{laa}     

\newcommand{\etal}{{\it et al.}}

\newcommand{\Om}{\Omega}
\newcommand{\de}{\delta}
\newcommand{\al}{\alpha}
\newcommand{\lam}{\lambda}
\newcommand{\La}{\Lambda}

\newcommand{\be}{\begin{equation}}
\newcommand{\ee}{\end{equation}}
\newcommand{\ra}{\rightarrow}
\newcommand{\bef}{\begin{figure}}
\newcommand{\eef}{\end{figure}}
\newcommand{\Mpc}{{\rm Mpc}}

\def\spose#1{\hbox to 0pt{#1\hss}} 
\def\ltapprox{\mathrel{\spose{\lower 3pt\hbox{$\mathchar"218$}} 
 \raise 2.0pt\hbox{$\mathchar"13C$}}} 
\def\gtapprox{\mathrel{\spose{\lower 3pt\hbox{$\mathchar"218$}} 
 \raise 2.0pt\hbox{$\mathchar"13E$}}} 
\def\inapprox{\mathrel{\spose{\lower 3pt\hbox{$\mathchar"218$}} 
 \raise 2.0pt\hbox{$\mathchar"232$}}} 
 
\begin{document} 
   \thesaurus{        
              (12.12.1;  
               12.04.1;  
               11.03.1)  
             }

\title {A Fractal  Galaxy Distribution in a Homogeneous Universe?}
\author {Ruth Durrer \inst{1} and Francesco Sylos Labini \inst{1,2}}
 \institute{        
                D\'ept.~de Physique Th\'eorique, Universit\'e de Gen\`eve,  
		24, Quai E. Ansermet, CH-1211 Gen\`eve, Switzerland.
		\and 
		   INFM Sezione Roma1,        
		      Dip. di Fisica, Universit\'a "La Sapienza", 
		      P.le A. Moro, 2,  
        	      I-00185 Roma, Italy. 
		} 
\date{Received -- -- --; accepted -- -- --} 

\maketitle
\begin{abstract} 
In this letter we present an idea which reconciles a homogeneous and
isotropic Friedmann universe with a fractal distribution of galaxies.
We use two observational facts: The flat rotation curves of galaxies
and the (still debated) fractal distribution of galaxies with fractal
dimension $D=2$. Our idea can also be interpreted as a redefinition of
the notion of bias.\\

\end{abstract}
              
 It is known since twenty years that the galaxy distribution
exhibits fractal behavior on small scales (Peebles 1980, Mandelbrot, 1982).
Several recent statistical analyses of three
dimensional galaxy catalogues indicate that galaxies are distributed
fractally with  dimension $ D\simeq 2$ out to the largest scales
for which statistically significant data is available,
 {\bf\it i.e.}, up to
about $  100$--$ 200h^{-1}$Mpc (Coleman \& Pietronero, 1992; Sylos
Labini \etal, 1998). In the opinion of the authors
of the present note, no  data is  pointing convincingly towards a
homogenization of the galaxy distribution ( {\bf\it i.e.}, $ D=3$ ). 
We believe that the standard analyses which indicate that fluctuations
of the galaxy number density decrease on large enough scales are
 not well suited to study scale invariant structures, since they 
a priori assume the existence of a well defined average density 
inside the given sample (Peebles, 1993; Davis 1997).  
On this point the scientific community has not reached
consensus (Wu \etal, 1998). In this work we assume without further
arguments that the fractal picture is correct. Our main point here is
to show that a fractal distribution of galaxies even up to the Hubble
scale may be consistent with a homogeneous and isotropic universe.

On the other hand, many observations, most notably the superb 
isotropy of the cosmic microwave background  together with its 
perfect blackbody spectrum, give strong evidence that the  
geometry of our Universe is very homogeneous and isotropic on large
enough scales, a so-called Friedmann model. 

In this letter we argue that this seemingly flagrant
contradiction of different pieces of observational data may actually
be reconciled in a rather simple way.

Our  main point is that a fractal distribution of
galaxies need not imply a fractal matter distribution (as it 
is also
pointed out in Wu \etal, 1998). 

It is often mentioned that  galaxies represent peaks in the matter 
distribution.
Let us compare these to the distribution of mountain peaks on the 
surface of the earth which we know  are fractally distributed over a
certain range of scales. But it would be  false to conclude
from this fact that the radius of the  surface of the earth from its center
is grossly variable; as we know, this number is very well
approximated by a constant. A similar effect may actually be at work
in the matter distribution of the universe. 

We first interpret the  COBE DMR
results (Bennett \etal, 1996) and other observations of CMB anisotropies on
smaller angular scales (De Bernardis \etal, 1997).
They indicate that matter 
fluctuations are reasonably well described by a Harrison-Zel'dovich spectrum 
with $(\de M/M)(\lambda_{\rm H}(t))\simeq 10^{-4}$,
where $\lambda_{\rm H}(t)$ denotes the comoving horizon scale,
$\lambda_{\rm H}(t)=\eta\sim t/a$, and $a$ is the scale factor.
But small density perturbations in a Friedmann 
universe grow only once the universe becomes matter dominated 
and even then rather
 slowly (proportional to the scale factor $a\propto t^{2/3}$). 
 Density fluctuations today on a given comoving scale  
$\lam$  should thus be on the order of
$\lam^{-3/2}\de(\lam)=10^{-4}(t_{\rm 0}/t(\lam))^{2/3} 
\sim 10^{-4}(\eta_{\rm 0}/\lam)^2$. 
Here $t(\lam)$ denotes the cosmic time at which the scale $\lam$ enters
the horizon; $t_0$ and $\eta_0$ are the present cosmic and conformal
times respectively. For example on the scale $\lam = 200h^{-1}$Mpc matter
density fluctuations should not be larger than 
\begin{eqnarray*}
 \left.\left({\de M\over M}\right)\right|_{\lambda=200h^{-1}\Mpc} &=&
	(\lambda^{-3/2}\de)|_{\lambda=200h^{-1}\Mpc} \\
&\sim& (3000h^{-1}{\rm Mpc}/200h^{-1}\Mpc)^2 10^{-4} \\
 &=& 0.015
 \end{eqnarray*}
 today.
(Here $h$ parameterizes the uncertainty of relating the recession velocity
of a galaxy to its distance, {\it i.e.} the uncertainty in the Hubble
parameter $H_{\rm 0}=100h {\rm km/s/Mpc}$.)

We now argue that a fractal galaxy distribution may well be compatible
with such a homogeneous Universe.
This can be seen by the following observation: It is well known that
the dark matter distribution around a galaxy  leads to
flat rotation curves (Rubin \etal, 1980).
 Without cutoffs this yields a matter density
distribution
\[  \rho_{\rm halo}(r) \propto r^{-2}  \]
around each galaxy. To obtain the total matter distribution, we
actually have to convolve $\rho_{\rm halo}$ with the number distribution
of galaxies, $n_{\rm G}$. The fractal dimension $D=2$ means that the average
number of galaxies in a sphere of radius $r$ around a given galaxy,
denoted by $N_{\rm G}(r)$, scales like $r^2$. This implies that the mean 
number  density of galaxies around an occupied point decays like 
\[ n_{\rm G} \propto {1\over r} ~.\]
Our main finding is the simple fact that the convolution of these 
two densities gives a constant (up to logarithmic corrections)
\be 
\rho = n_{\rm G}*\rho_{\rm halo} \propto1~~.  
\label{conv}
\ee
More precisely, for $n_{\rm G}(r) = C/r$ and 
$\rho_{\rm halo}(r)=A/r^2$ we
obtain ( with $ |{\bf y}|\equiv y$ and  $ |{\bf x}|\equiv x$)
\begin{eqnarray}
 \rho({\bf x}) =AC\int d^3y {1\over |{\bf y}| |{\bf x-y}|^2}
	= 4\pi AC \int_0^{R_{\rm max}}{\min(x,y)\over xy}dy
\nonumber   \\
 = 4\pi AC\left[{1\over x}\int_0^x dy + 
\int_x^{R_{\rm max}}{dy\over y}\right]
        = 4\pi AC[1+ \ln(R_{\rm max}/x)] ~.
\label{rho}
\end{eqnarray}
This shows that a  fractal galaxy distribution with dimension $D=2$
together with flat rotation curves, indicate a
smooth matter distribution in agreement with our expectations from a
Friedmann universe with small fluctuations. The dark matter
distribution of a two dimensional model where the galaxies are
distributed with fractal dimension $D=1$ is shown in
Fig.~\ref{fig1}. Clearly, this dark matter distribution is very
homogeneous up to finite size effects.

Note that the essential ingredient for this result is that
  $n_{\rm G} \propto 1/r^\al$ and  $\rho_{\rm halo}(r)\propto 1/r^\beta$
with $\al+\beta =3$.

\bef 
\vspace{6cm}
\caption{ \label{fig1} 
We show the dark matter distribution of a two dimensional set of
fractally distributed galaxies 
$(D=1, n_{\rm G} \propto 1/r)$ (filled circles) 
each surrounded by a dark matter  (dots) halo with distribution
$\rho_{\rm halo}\propto 1/r$. The halos sum up to a very
homogeneous dark matter distribution.}
\eef 

It is important to insist that the cutoff $R_{\rm max}$ is larger
than all the scales $x$ considered. Far away from galaxies, for
example in a void, many galaxies contribute to the density in a given
point and the important message is just that flat rotation curves
indicate actually that the total resulting density may be rather
constant in voids and has about the same value as it has close to galaxies. 

One may argue that the notion
of a certain lump of matter 'belonging' to a certain galaxy is ill
defined sufficiently far away from a galaxy and thus the 'halo'
density cannot be defined on scales larger than, say, half the distance to
the next galaxy. With this objection we agree in practice, it just
means that at a sufficient distance from a given galaxy the only
measurable density is the total density which contains relevant
contributions from many galaxies.  But in
principle it is possible to assign to each galaxy a density profile 
$\rho_{\rm halo}$, and  it is interesting to note that the form of
the density profile indicated by measurements, which are
possible close to isolated galaxies, is just such as to lead to a
constant total density if we convolve it with the number
density of galaxies.

Another objection may be, that with this density profile the total mass of a
galaxy is infinite. But this is in fact irrelevant, since not only
the mass of a single galaxy goes to infinity as $r \ra \infty$, but also
the galaxy density goes to zero as $r \ra \infty$ and this just in a
way that the measurable total density is constant. Besides, the
integral of $\rho_{\rm halo}$  
should not be considered as the 'mass of the
galaxy'. More profoundly, the flat rotation curves are a consequence
of the fact that the gravitational potential remains constant during linear
clustering in a Friedmann universe, which holds beyond the scale of
single galaxies.

Clearly, the amplitudes $C$ and $A$ of 
$n_G$ and $\rho_{\rm halo}$ depend
on the type of galaxies considered. Galaxies of different absolute 
luminosities in general have different circular speeds and different
abundances. 

To quantify Eq.~(\ref{rho}) we use the Tully-Fisher relation (Tully
\& Fisher, 1977), between
 the circular speed $v$ of a galaxy and its luminosity $L$,
 $v=v_*(L/L_*)^{0.25}$, with $v_* =220km/s$. The luminosity
$L_*$ corresponds to an absolute magnitude  of $M_* \simeq -19.5$.

The abundance of $L_*$ galaxies as estimated from Sylos
Labini \etal (1998) is
\[ N_{\rm G}(r)_* =B_*r^2  
~~~\mbox{ with }~~~ B_* \simeq {0.3h^2\over {\rm Mpc}^2}
~.\]
Considering for the time being just $L_*$ galaxies, this
gives $n_{\rm G}=C_*/r$ with $C_*\simeq 3B_*/4\pi$.

To determine the amplitude of the halo density, 
$\rho_{\rm halo} =A/r^2$,
we use the flatness of galaxy rotation curves with (Rubin \etal, 1976)
\[ {1\over 2}(v_*/c)^2 \simeq 10^{-7}=GM(r)/r ~,\]
where $c$ denotes the speed of light.
Combining this with $M(r) = 4\pi Ar$ gives
\[ A_* = {(v_*/c)^2\over 8\pi G} ~.\]
The convolution of $n_{\rm G}$ 
with $\rho_{\rm halo}$ then leads to
\be
\rho = n_{\rm G} * \rho_{\rm halo} 
\simeq 4\pi C_*A_*\simeq {3B_*(v_*/c)^2\over 8\pi G} ~.
\ee
Comparing this with the critical
density of a universe expanding with Hubble parameter $H_0$,
\be 
	\rho_c = {3H_{\rm 0}^2c^2\over 8\pi G} ~,
\ee
we obtain a density parameter of order unity, 
\be \Om =\rho/\rho_c \simeq v_*^2B_*/H_0^2 \sim 1~. \label{den}\ee 
Clearly, this estimate is very crude since not all galaxies have the
same rotation speeds and the constant $B$ depends on the 
luminosity of the galaxy. But it is a  reassuring non-trivial
'coincidence' that the density parameter obtained in this way is of the
right order of magnitude. To refine this model we have to find a
Tully-Fisher type relation for $B(L)$ and integrate over luminosities.
But since the abundance of galaxies decays exponentially with
luminosity above $L_*$, we have
\[ \int v(L)^2dB(L) \sim v_*^2B_* ~,\]
\noindent and reproduce the result (\ref{den}).

Our arguments suggest that a fractal galaxy distribution may well be
in agreement with a smooth matter distribution. Neglecting log
corrections (which may be absent in a more realistic, detailed model
and which are certainly not measurable with present accuracy), our
model describes a perfectly homogeneous and isotropic universe. We do
not specify the process which has induced small initial fluctuations
and finally led to the formation of galaxies. We are thus still
lacking a specific picture of how the fractal distribution of 
galaxies may have emerged. Purely Gaussian initial fluctuations are
probably not suited to reproduce a fractal galaxy distribution. But
cosmic strings or other 'seeds' with long range correlation could do
it. A working model remains to be worked out.

It may be useful to mention that the view presented here actually
redefines the notion of bias.  In the standard  scenario, the 
 dark matter density field (on large scales) is Gaussian and galaxies 
form in the peaks of the underlying distribution and their correlation
functions are simply related\cite{Kaiser}
Here we consider the possibility that the
galaxy and dark matter distributions may have different correlation
properties and hence different fractal dimensions, namely $D=2$ for
the galaxies and $D=3$ for dark matter. Especially in view of 
Eq.~(\ref{conv}), we want to warn the reader against interpreting 
the galaxy distribution as proportional to the
matter distribution even on large scales. 

A similar idea is the one of a universe with a fractal galaxy
distribution but a dominant cosmological constant $\La> 8\pi G\rho$. This
 possibility is also discussed in Baryshev \etal (1998),
in connection with the linearity of the Hubble law.

More precisely, the fractal 
dimension of a set of density fluctuations can depend on the 
threshold (see Fig.~\ref{fig2}). In a realistic model, we would expect
that also the dark matter, above a certain threshold is fractally
distributed. In galaxy catalogues, this tendency is actually 
indicated. Observations  show a slight increase of the fractal 
dimension with decreasing
absolute luminosity of the galaxies in the sample (Sylos Labini \etal, 1998), 
however, still with relatively modest statistics. 
Such a scenario is naturally
formulated within the framework of multi-fractals
(Falconer, 1990;  Sylos Labini \etal, 1998). It is, for example, a
well known fact that bright ellipticals lie preferentially in clusters
whereas spiral galaxies prefer the field (Dressler, 1984). From the
perspective of multi-fractality this implies that ellipticals are more
clustered than spirals, {\em i.e.}, their fractal dimension is 
lower (Giovanelli \etal, 1986). 

\bef 
\epsfxsize 10cm 
\centerline{\epsfbox{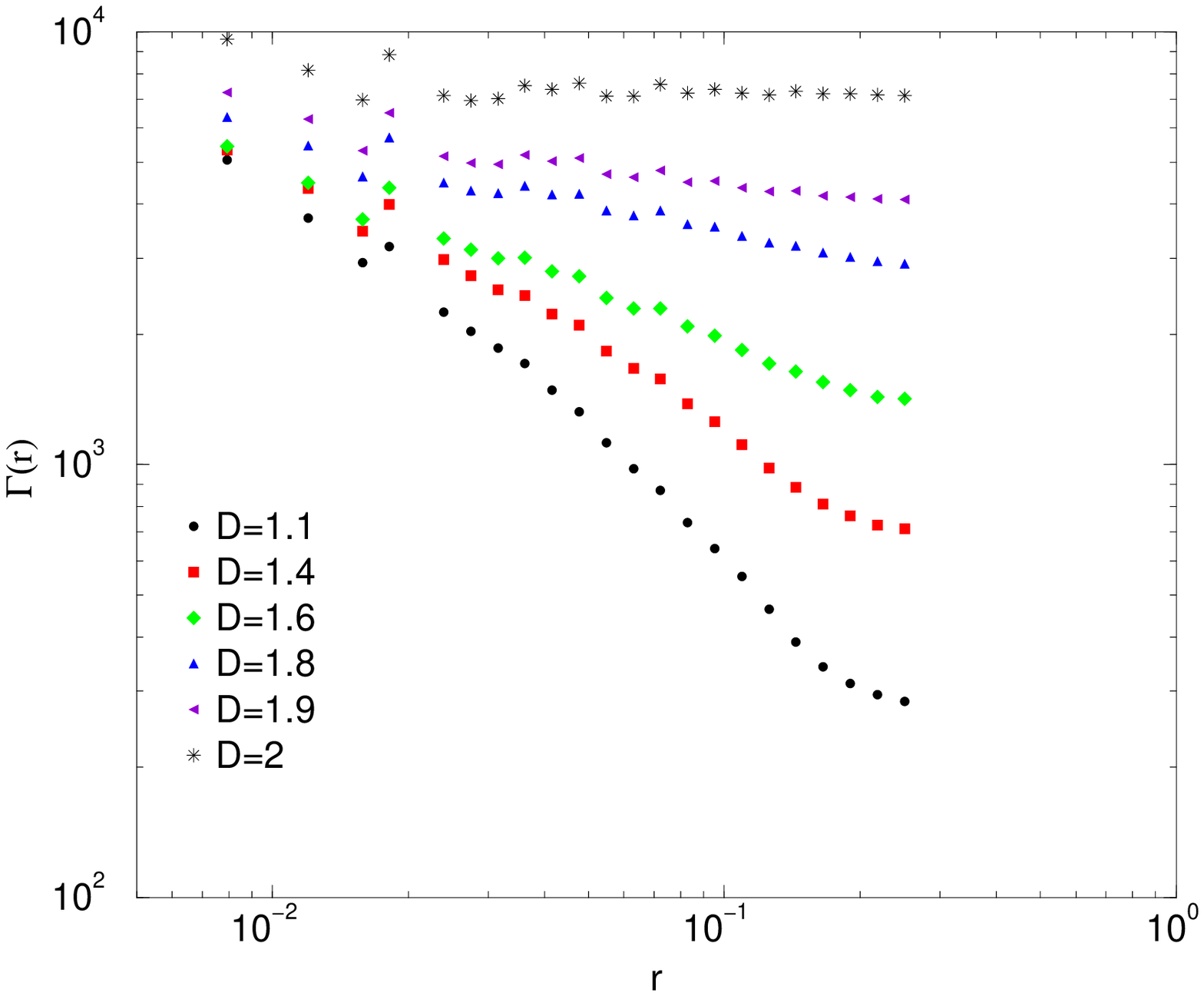}} 
\caption{ \label{fig2} 
The mean density around an occupied point $(\Gamma(r))$ is shown for
our two dimensional model. The galaxy density with fractal dimension $D=1$
is shown as filled circles. As the density threshold  decreases the 
fractal dimension approaches $D=2$. The chosen overdensities for the
fractal demensions of 1.1, 1.4, 1.6, 1.8, 1.9 and 2 are 2.1, 1.76, 1.55,
1.34, 1.22 and 1 respectively.
} 
\eef 
Our arguments indicate that the voids might be filled by
dark matter. But since this dark matter is relatively smoothly
distributed, it cannot be detected by measurements sensible only to 
density gradients (like, {\em e.g.}, peculiar velocities).
One needs to determine
the total density of the universe, for example by measuring the
deceleration parameter $q_{\rm 0}$. 
\vspace{0.1cm}\\

{\bf Acknowledgment:} It is a pleasure to thank Yuri Baryshev, 
Alessandro Melchiorri,
Luciano Pietronero, Pekka Teerikorpi
and Filippo Vernizzi for valuable discussions and comments.
This work has been partially supported by the 
EEC TMR Network  "Fractal structures and  self-organization"  
\mbox{ERBFMRXCT980183} and by the Swiss NSF.

\end{document}